\documentclass[article,shortnames]{jss}\usepackage[]{graphicx}\usepackage[]{color}
\pdfoutput=1
\makeatletter
\def\maxwidth{ %
  \ifdim\Gin@nat@width>\linewidth
    \linewidth
  \else
    \Gin@nat@width
  \fi
}
\makeatother

\usepackage{Sweave}

\usepackage{thumbpdf,lmodern}

\usepackage{bm}

\usepackage{pdflscape}
\usepackage{subfig}

\usepackage{amsmath}

\newcommand{\rbest}{\pkg{RBesT}}

\newcommand{\normal}{\mbox{Normal}}
\newcommand{\hnormal}{\mbox{HalfNormal}}
\newcommand{\dbeta}{\mbox{Beta}}
\newcommand{\dgamma}{\mbox{Gamma}}
\newcommand{\logit}{\mbox{logit}}
\newcommand{\betabinomial}{\mbox{BetaBinomial}}
\newcommand{\dbinomial}{\mbox{Binomial}}
\newcommand{\dexp}{\mbox{Exp}}
\newcommand{\cp}{\mbox{CP}}
\newcommand{\pos}{\mbox{PoS}}
\DeclareMathOperator*{\argmax}{arg\,max}

\newcommand{\pd}[3][]{ \partial^{#1}_{#3} #2}

\newcommand{\dpois}{\mbox{Poisson}}

\makeatletter
\newcommand{\eg}{e.g\@ifnextchar.{}{.\@}}
\makeatother

\makeatletter
\newcommand{\ie}{i.e\@ifnextchar.{}{.\@}}
\makeatother

\usepackage[utf8]{inputenc}

\author{Sebastian Weber\\Novartis Pharma AG
   \And Yue Li\\Novartis Pharma AG
   \And John W Seaman III\\Novartis Pharma AG \AND
   \And Tomoyuki Kakizume\\Novartis Pharma K.K.
   \And Heinz Schmidli\\Novartis Pharma AG}
\Plainauthor{Sebastian Weber, Yue Li, John Seaman, Tomoyuki Kakizume, Heinz Schmidli}

\title{Applying Meta-Analytic-Predictive Priors with the \proglang{R} Bayesian evidence synthesis tools}
\Plaintitle{Applying Meta-Analytic-Predictive Priors with the R Bayesian evidence synthesis tools}
\Shorttitle{Applying MAP Priors with \rbest }

\Abstract{ Use of historical data in clinical trial design and
  analysis has shown various advantages such as reduction of
  within-study placebo-treated number of subjects and increase of
  study power. The meta-analytic-predictive (MAP) approach accounts
  with a hierarchical model for between-trial heterogeneity in order
  to derive an informative prior from historical (often control)
  data. In this paper, we introduce the package \rbest\ (\proglang{R}
  Bayesian Evidence Synthesis Tools) which implements the MAP approach
  with normal (known sampling standard deviation), binomial and
  Poisson endpoints. The hierarchical MAP model is evaluated by
  MCMC. The numerical MCMC samples representing the MAP prior are
  approximated with parametric mixture densities which are obtained
  with the expectation maximization algorithm. The parametric mixture
  density representation facilitates easy communication of the MAP
  prior and enables via fast and accurate analytical procedures to
  evaluate properties of trial designs with informative MAP
  priors. The paper first introduces the framework of robust Bayesian
  evidence synthesis in this setting and then explains how \rbest\
  facilitates the derivation and evaluation of an informative MAP
  prior from historical control data. In addition we describe how the
  meta-analytic framework relates to further applications including
  probability of success calculations.}

\Plainkeywords{Bayesian inference, clinical trial, extrapolation,
  historical control, operating characteristics, prior, probability of
success, robust analysis}

\Address{
  Novartis Pharma AG
  4002 Basel, Switzerland\\
  E-mail: \email{sebastian.weber@novartis.com}
}
\IfFileExists{upquote.sty}{\usepackage{upquote}}{}
\begin{document}





\section{Introduction}
\label{sec:intro}

More efficient clinical trials are of great demand in drug development
for all players like pharmaceutical companies, regulatory agencies,
health-care organizations and, most importantly, for patients. Use of
historical data for quantitative trial design has become more and more
attractive for the same reason \citep{wan2016uhd,
  neuenschwander2019}. Using historical data can reduce the size of
the control group, leading to a smaller size of clinical trials which
are more ethical and shorten study duration. Therefore, studies
utilizing historical data may speed up informative decision making and
eventually make better medicines available to patients sooner.

Borrowing information from historical studies has always been a part
of the design of clinical trials. For example, the definition of a
patient population in a new clinical study compared to previous
similar studies, or how much of a clinically relevant treatment effect
to expect compared to the control treatment. Contributions from
statisticians in a more quantitative manner started more than 40 years
ago, by \cite{poc1976crh}. Since then, the relevant statistical
approaches have been developed by many, mostly in a Bayesian framework
\citep{che2000ppd, spi2004bac, neu2010shi, hob2012cpi}.

In this paper, we focus on robust meta-analytic-predictive (MAP)
priors \citep{neu2010shi, spi2004bac}, which summarize the historical
information as an informative prior to be used in a Bayesian analysis
for the study of interest. The MAP prior is derived using a random
effects meta-analysis model, also referred to as hierarchical
model. The meta-analysis accounts for between-trial heterogeniety and
leads to a discounting of the historical information. As in any
meta-analytic approach, it is important to first examine the
characteristics of the historical trials with clinical inputs. These
include quantitative descriptions of trial population such as subject
demographics, baseline characteristics and qualitative features such
as concomitant medications. This helps to ensure that selected
historical controls are comparable to those in the new trial, such
that one of the key assumptions of exchangeability between the trials
holds. Secondly, the assumption for the between-trial heterogeneity
needs to be reasonable as the number of trials is commonly small to
borrow from and considerable uncertainty needs to be accounted for
adequatley in the analysis. Then the MAP prior can be derived from a
random-effect meta-analysis of historical data via Markov Chian Monte
Carlo (MCMC) algorithms as the analysis is commonly not tractable
analytically. The predictive distribution of the effect parameter for
a new study then forms the informative prior for the study of
interest. In many cases, the goal is to replace some sample size of
the planned control group with information from the MAP prior. The
effective sample size of the MAP prior can then be used as a
convenient measure of informativeness for the potential sample
size savings of the actual trial. However, since in practice we can
never rule out the unexpected, a robustification of the MAP prior
\citep{sch2014rmp} is advisable in order to ensure sensible inference
whenever actual trial data and MAP prior strongly deviate from one
another. Application of the MAP approach to incorporate historical
data in early phase clinical trials has been more widely accepted, not
only by statisticians but also in medical societies
\citep{Baeten2013}.

While there exists a number of \proglang{R} packages which perform
meta-analysis, these target the conduct of the meta-analysis for the
purpose of combining the available historical information
alone. However, these lack the aspect of using the combined historical
information for the purpose of a trial design and analysis. For
example, \pkg{netmeta} implements a frequentist approach
\citep{ruecker2012}, \pkg{bayesmeta} for random-effect meta-analysis
\citep{bayesmeta}, \pkg{metafor} for mixed-effect meta-analysis and
meta-regression is provided by \citep{metafor}, and \pkg{MetaStan}
\citep{Gunhan2019}. There are also \proglang{R} packages developed for
network meta-analysis, which seeks to combine the historical
information for multiple treatments, \ie\ \pkg{gemtc} uses JAGS for
arm-based network meta-analysis \citep{gemtc}, \pkg{pcnetmeta} is
based on contrast-based network meta-analysis \citep{pcnetmeta} and
\pkg{nmaINLA} employs the integrated nested Laplace approximations
\citep{nmaINLA}.

In this \proglang{R} package, \proglang{R} Bayesian evidence synthesis
tools, \pkg{RBesT}, we implement the MAP approach tailored for the
design and analysis of clinical trials as conducted during (usually
early phases of) drug development. The focus on trial design leads to
a number of considerations specific to this application. For example,
the limited sample sizes involved may not warrant that the often used
normal approximation holds well enough such that in addition to the
common normal endpoint a binomial and a Poisson endpoint are supported
by \pkg{RBesT}. As the package facilitates trial design and clinical
trial protocol writing, a key step, which makes \pkg{RBesT} unique, is
that the derived MAP prior can be easily documented in parametric form
and, most importantly, the implications of actually using an
informative prior for the trial design can be readily evaluated. Thus,
in \pkg{RBesT} the meta-analysis is the starting point and large
emphasis is on facilitating trial design and supporting compilation of
the statistical analysis plan which pre-specifies the (primary) trial
analysis.

In order to document and easily pre-specify an informative MAP prior
for a trial, \pkg{RBesT} offers functionality to approximate the
posterior MCMC sample representing the MAP prior with a parametric
mixture distribution. The parametric mixture distribution is
arbitrarily exact and is estimated using the expectation maximization
(EM) algorithm. Representing the MAP prior in parametric form is key
to design clinical trials. For one, the exact statistical analysis of
the trial results are pre-specified ahead of time of trial analysis
and, for two, the parametric representation allows seamless evaluation
of the design in question. For each endpoint specific conjugate
mixture densities are estimated. This allows the use of accurate and
fast analytical results. \pkg{RBesT} supports trial design evaluation
of one or two sample situations with flexible success criterions. In
addition to standard frequentist operating characteristics, the
probability of success or derivation of critical values is
supported. All of these subsequent steps after MAP prior derivation
operate in \pkg{RBesT} on the parametric mixture densities such that a
large fraction of the package is devoted to evaluation of mixture
densities.


The organization of this paper is as follows. We will first give an
overview of the package key functionality and describe the details of
one motivating example from a real clinical study. In Section 3,
theoretical background on MAP approach and robustification of MAP will
be explained, followed by how these approaches can be applied via
\pkg{RBesT} to the example. Finally we close the paper with summary
and discussions.



\section{Historical control in Phase II}
\label{sec:example}

One of the main use-cases of \rbest\ is the use of historical
information in the analysis of clinical trials. The goal is to reduce
the trial (usually the control group) sample size while maintaining
its target power under the assumed true effect size of an alternative
hypothesis. The \rbest\ package facilitates the (i) prior derivation
using MCMC, (ii) parametric (mixture) approximation of the MAP prior
and finally (iii) evaluation of the clinical trial design.

\paragraph{Prior derivation} Once relevant data for the prior
derivation has been compiled, the first step is to use the \code{gMAP}
function which performs a random-effects meta-analysis using MCMC
implemented by \pkg{rstan} \citep{RStan}. The syntax follows standard
\proglang{R} modeling conventions and the returned analysis object can
be modified for sensitivity analyses with the \code{update} command or
queried with the generic modeling functions. In particular, the
\code{plot} function creates graphical model diagnostics like a model
estimate augmented forest plot. As the forest plot is a key summary of
the data making up the prior a dedicated \code{forest_plot}
function is available which allows for various customization of the
forest plot.

\paragraph{Parametric mixture approximation} After running the MCMC
analysis with \code{gMAP}, the next step is to find an accurate
parametric distributional representation. The package uses the EM
algorithm implemented by the \code{mixfit} function which can be
applied to \code{gMAP} analysis objects or MCMC samples directly. The
EM procedure requires specification of the number of mixture
components to be used. To ease this step even further, the package
also provides an \code{automixfit} function which selects the number
of mixture components automatically based on information criteria
measures. The resulting parametric mixture distribution objects
generated from \code{mixfit} also support the \code{plot} method which
produces a graphical check of the fitted mixture
distribution by a comparison to an histogram of the MCMC sample.

\paragraph{Trial design evaluation} It is first recommended to
robustify the parametric MAP prior in with the \code{robustify}
function. This function adds a weakly-informative prior component to
the mixture derived from the previous steps. The so-obtained robust
MAP prior is intended for use in a pre-specified trial analysis. It's
informativeness can be assessed with the effective sample size with
the \code{ess} command. At the design stage of the trial the
frequentist operating characteristics or probability of success of the
trial are of interest in order to document the properties of the
statistical analysis as planned in the statistical analysis plan of
the trial. The trial evaluation is supported in \rbest\ for one- or
two-sample designs. In a first step, the statistical success criterion
of the trial is specified with the \code{decision1S}
(\code{decision2S}) function which return objects representing the
decision function. These decision functions are then required by the
functions \code{oc1S} (\code{oc2S}), \code{pos1S} (\code{pos2S}) and
\code{decision1S_boundary} (\code{decision2S_boundary}) to specify the
trial design (defined by prior, sample size and decision function) for
the purpose of evaluating the frequentist operating characteristics,
the probability of success or the critical decision boundaries of the
trial design, respectively.

\paragraph{Ankylosing spondylitis Phase II example}
As an example we will use a Novartis Phase II study in ankylosing
spondylitis comparing the Novartis test treatment secukinumab with
placebo \citep{Baeten2013}. The primary efficacy endpoint was a binary
responder analysis for the percentage of patients with a 20\% response
according to the Assessment of SpondyloArthritis international Society
criteria for improvement (ASAS20) at week 6. Eight historical trials,
totaling $533$ patients as shown in table~\ref{tab:as}, were used to
derive the MAP prior for the control arm.

\begin{table}[h]
\centering
\begin{tabular}{lcccccccc}
\hline
Study & 1   & 2  & 3  & 4  & 5   & 6  & 7  & 8  \\ \hline
Patients (n)     & 107 & 44 & 51 & 39 & 139 & 20 & 78 & 35 \\
Responders (r)     & 23  & 12 & 19 & 9  & 39  & 6  & 9  & 10 \\ \hline
\end{tabular}
\caption{Historical data used in Novartis Phase II study in ankylosing
  spondylitis. The data set is available in the \code{AS} data frame
  as part of the \rbest\ package.}
\label{tab:as}
\end{table}

This Novartis Phase II trial was conducted using the MAP approach
(before the availability of \rbest). The trial used a $\dbeta(11,32)$
prior for the control arm and performed a $4:1$ randomization ratio of
active vs control patients. The final trial compared $n=24$ ($r=15$)
treated vs $n=6$ ($r=1$) control patients and declared success based
on meeting the success criterion defined as requiring that
$P(\delta \leq 0|y) > 0.95$ holds. This example is discussed with
greater focus on the statistical aspects in the vignette ``\rbest\ for
a Binary Endpoint'' part of \rbest.

\section{Bayesian evidence synthesis and prediction}
\label{sec:BES}

Important decisions should arguably be evidence based, especially in
medicine \citep{eddy1990, wan2016dac}.  For example, decisions
regarding design and analysis of clinical trials are important for
trial sponsors, patients, physicians and policy makers.  To
support such decisions, relevant sources of information should be
collected, appropriately synthesized through meta-analytic approaches,
and used to make predictions on the planned target trial.  We use the
term meta-analytic-predictive (MAP) approach \citep{neu2010shi} to
denote the synthesis of evidence from various sources, and the
prediction/extrapolation to the target. Although the MAP approach is
useful in a broad range of applications, we consider here the medical
setting \citep{ema2013extrap}.

Methodology for Bayesian evidence synthesis and prediction is well
developed. Textbooks on this topic include \cite{stangl2000meta},
\cite{spi2004bac}, \cite{welton2012evidence}, and
\cite{dias2018network}. Robust hierarchical models play a key role
here, as explained in the following Sections \ref{sec:BESH} and
\ref{sec:BESR}.  Applications of the MAP approach are very diverse,
and we briefly discuss some settings in Section \ref{sec:BESA}.

\subsection{Meta-analytic-predictive methodology}
\label{sec:BESH}

\begin{figure}[t!]
\centering
\includegraphics[scale=.08]{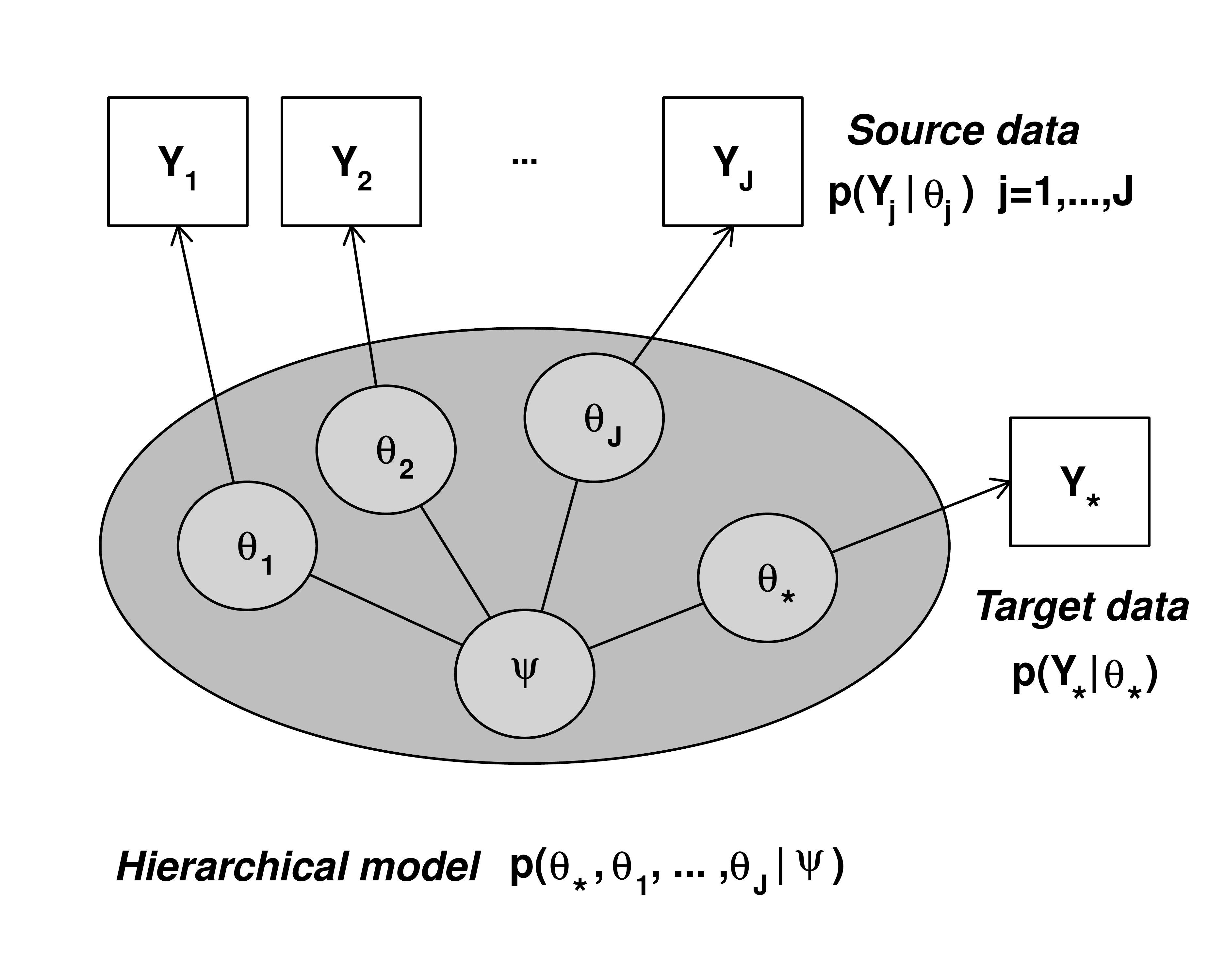}
\caption{\label{fig:ES} MAP approach to evidence synthesis and prediction.}
\end{figure}

Figure~\ref{fig:ES} schematically depicts the MAP approach for
evidence synthesis and prediction.  Suppose that a sponsor plans a new
clinical trial (the target, labeled by the star symbol).  This trial
will generate data $Y_\star$, to be described by a statistical model
$p(Y_\star | \theta_\star)$, with parameters $\theta_\star$.  Usually,
several relevant sources of information will be available,
\eg\ clinical trials in the same or similar patient population, and
with (partly) the same treatments.  Each source of information
consists of data $Y_j$, modeled by $p(Y_j | \theta_j)$, with
corresponding parameters $\theta_j$, $j=1,...,J$.  To borrow strength
from the source information, a model is required that links the
parameters from both source and target:
$p(\theta_\star, \theta_1, ..., \theta_J | \Psi )$, with
hyper-parameters $\Psi$.  Such hierarchical models are very natural
and convenient for the synthesis of the evidence and the prediction to
the target \citep{spi2004bac}.  Within the Bayesian framework, a prior
for the hyper-parameters $p(\Psi)$ is needed, which will be specific
to the considered setting.

At the planning stage of the target trial, the data $Y_\star$ are not
available. Hence the posterior distribution of the parameters
$p(\theta_\star, \theta_1, ..., \theta_J, \Psi | Y_1, ..., Y_J )$ is
based on the source data only. The marginal posterior for the target
parameter $p_{MAP}(\theta_\star) = p(\theta_\star | Y_1, ..., Y_J )$
is the prior information for the target, called the MAP prior in the
following.

Once the target data $Y_\star$ are available, the posterior for
$\theta_\star$ is
$p(\theta_\star | Y_\star) \propto p(Y_\star | \theta_\star)
p_{MAP}(\theta_\star)$.  Exactly the same posterior could also be
obtained through a meta-analytic-combined (MAC) approach, from
$p(\theta_\star, \theta_1, ..., \theta_J, \Psi | Y_\star, Y_1, ...,
Y_J )$ \citep{sch2014rmp}.

An analytical derivation of the MAP prior is typically not possible,
and hence Markov chain Monte Carlo (MCMC) methods have to be used
\citep{gelman2013bayesian}. These generate a large sample from the
posterior distribution of the parameters, including $\theta_\star$.
However, an approximate analytical description of the MAP prior
facilitates communication and use with standard software.  Mixtures of
normal distributions generally provide such an analytical
approximation \citep{west1993approximating}.  When conjugate priors
exist, mixtures of conjugate priors may be used \citep{dal1983apb,
  dia1984qpo}.  These are preferable, as they allow analytical
posterior calculations for the target trial in simple settings
\citep{oha2004bi, sch2014rmp}.

\subsection{Robustness to prior-data conflict}
\label{sec:BESR}

The MAP approach provides the prior for the target parameters as
$p_{MAP}(\theta_\star)$.  Occasionally, this MAP prior may turn out to
be in conflict with the emerging target data $Y_\star$, despite great
care in the selection of relevant sources and in the specification of
the model.  The behavior of the posterior distribution to prior-data
conflicts is governed by the tails of the prior \citep{oha2012bhm}.
For example, conjugate priors are not heavy-tailed, and consequently
the posterior will always be a compromise between prior and data.
However, MAP priors are typically heavy-tailed, and are essentially
discarded in case of prior-data conflict. This is a desirable feature
in most settings.

Although the MAP prior should be robust to prior-data conflict by
accounting for heterogeneity, a faster reaction to prior-data
conflicts may be achieved by adding a weakly-informative mixture
component $p_V(\theta_\star)$ \citep{sch2014rmp,
  neuenschwander2016ps}. The robustified MAP prior is:
\begin{equation} 
\label{eqn:rob}
p_{MAPr} (\theta_\star) = (1-w)\,p_{MAP}(\theta_\star) + w\,p_V(\theta_\star), 
\end{equation}
where $w$ may be interpreted as the prior probability that the
source information is not relevant for the target, expressing some
degree of skepticism towards borrowing strength.

\subsection{Applications}
\label{sec:BESA}

Methodology and diverse applications of the MAP approach in medicine
are reviewed in \cite{wan2016uhd}, \cite{schmidli2019}, and
\cite{neuenschwander2019}. Specific applications include comparison of
several treatments though a network MAP approach \citep{sch2013nma},
the design and analysis of non-inferiority and biosimilar clinical
trials \citep{gam2016bmd, mielke2018BJ}, and the use of external data
in adaptive clinical trials \citep{gsp2014pgt, mutze2018PS}. In the
following, four common applications are briefly described.

\subsubsection{Random-effects meta-analysis}
\label{sec:BESA.MA}

Random-effects meta-analyses of clinical trials are very common in
medicine \citep{hig2008chs}.  These typically synthesize the evidence
on the comparative effectiveness of two interventions in patients with
a specific disease.  \cite{hig2009rre} emphasize that both the overall
effect size and the prediction for the true effect in a new trial are
important for decision makers.

Sources of information are often $J$ randomized clinical trials
comparing a test and a control treatment, with a continuous clinical
endpoint. Data available from the $j$-th source trial are taken from
publications, and are usually the estimated effect $Y_j$ with standard
error $se_j$ (taken as exactly known). These data are modeled as
$Y_j \sim \normal(\theta_j, se_j^2)$.  The parameters are linked
through a model:
$\theta_\star, \theta_1, ..., \theta_J \sim \normal(\mu,\tau^2)$, with
overall effect size $\mu$ and between-trial standard deviation
$\tau$. The parameter $\theta_\star$ denotes the true effect in a new
trial.  Priors for the hyper-parameters $\Psi=(\mu,\tau)$ are \eg\ a
weakly informative Normal prior for $\mu$ and a Half-Normal prior for
$\tau$.  In case with few trials (\ie\ $J<5$), an appropriate choice
of the prior for $\tau$ is crucial
\citep{gel2006pdv,fri2017ma2,friede2017rsm}. After having specified
the priors, \rbest\ may be used to obtain a sample from the posterior
distribution of the parameters, and to graphically and numerically
summarize these.

\subsubsection{Evaluation of probability of success}
\label{sec:BESA.POS}

Clinical trials aiming to show the superiority of a test treatment
over a control treatment are often analyzed using a frequentist
approach.  The trial is considered a success, if a statistically
significant treatment effect is observed, with one-sided significance
level $\alpha=0.025$.  The sample size of the trial is chosen such
that a power of \eg\ 80\% is achieved, conditional on a specific
treatment effect $\theta_\star$.  However, the power does not
provide the unconditional probability of success (or assurance), as it
ignores the uncertainty on the treatment effect.  If relevant source
data on the treatment effect are available, the uncertainty on
$\theta_\star$ is captured by the MAP prior $p_{MAP}(\theta_\star)$.

The probability of success (PoS) is the prior expectation of the
power, averaged over the MAP prior \citep{o2005assurance}.
\begin{equation} 
\label{eqn:pos}
\pos = \int \cp(\theta_\star)  \, p_{MAP}(\theta_\star) \, d \theta_\star ~ , 
\end{equation}
where $\cp(\theta_\star)$ is the conditional power function, \ie\ the
probability of success conditional on an assumed true treatment
effect.  From a MCMC sample
$\theta_\star^{(1)}, ..., \theta_\star^{(M)}$ of the MAP prior the PoS
may be calculated as $1/M \sum_m \cp(\theta_\star^{(m)})$.
Alternatively, the MAP prior may be approximated by a mixture of
normal priors with \rbest, and PoS can be evaluated by numerical
integration.

PoS evaluations are also relevant for decision makers at interim
analyses of clinical trials. If the PoS (or predictive power) at
interim is low, the trial may be stopped early to avoid unnecessary
exposure of patients to ineffective treatments and to save resources
\citep{spiegelhalter1986cct, sch2007bpp, neu2016ucd}.  For these
interim analyses, the MAP prior is updated with the interim data.

\subsubsection{Extrapolation}
\label{sec:BESA.E}

New treatments are typically first investigated in adult patients,
before starting clinical trials in children.  For many diseases and
treatments, a similar effect may be expected for children and adults,
using a possibly modified children version of the treatment
(\eg\ dosing based on body weight).  Borrowing strength from the
available adult trials should therefore always be considered
\citep{fda2015lec, ema2018ped}.  The MAP approach may be used for
extrapolation \citep{wandel2017, roever2019}, although alternative
methods are also available \citep{gamalo2017pharmstat,
  wadsworth2018smmr}.

The source data are usually $J$ randomized clinical trials in adults
comparing test and control treatment.  These can be summarized with a
random-effects meta-analysis as described above, which provides the
MAP prior for the treatment effect in a new trial in adults
$p_{MAP}(\theta_\star)$.  In some settings essentially the same
treatment effect in adults and children may be expected, based on a
scientific understanding of the disease and the mode-of-action of the
treatment. Hence, the MAP prior for adults may also be used for a new
trial in children.  Skepticism on the relevance of the adult data may
be expressed by robustifying the MAP prior (Section \ref{sec:BESR}).
In simple settings, \rbest\ can be used to derive the MAP prior,
robustify it, evaluate frequentist operating characteristics of the
trial in children, and finally obtain the posterior distribution of
the treatment effect in children, once results from the children trial
are available.

\subsubsection{Historical controls}
\label{sec:BESA.HC}

In many disease areas, multiple randomized controlled trial (RCT) have
been conducted, with the same control group (\eg\ placebo) but
different test treatments.  When planning to investigate a new test
treatment in a RCT, the question arised whether one could borrow
strength from the historical control data \citep{vie2014uhc}.  In this
setting, the sources of information are the control data from $J$
trials. For a clinical endpoint, the control mean from the
$j$th trial may be modeled as $Y_j \sim \normal(\theta_j, se_j^2)$,
with true control mean $\theta_j$, and standard error $se_j$ (taken as
exactly known).  A model is used to provide the link to the true
control mean in the new trial $\theta_\star$ as:
$\theta_\star, \theta_1, ..., \theta_J \sim \normal(\mu,\tau^2)$.  With
appropriate priors for the hyper-parameters, the MAP prior
$p_{MAP}(\theta_\star)$ is derived, and used as the informative prior
for the control group in the new trial. Again, it is often advisable
to robustify the MAP prior in case of some doubt on the relevance of
the historical control information (Section~\ref{sec:BESR}). \rbest\ may
be used for MAP prior derivation, evaluation of frequentist operating
characteristics and the final analysis. An example data set is
described in Section~\ref{sec:example} and in
Section~\ref{sec:application} we present how \rbest\ facilitates the
use of historical control information in clinical trials.

Use of the MAP approach in historical control settings has also been
described for data modeled by the one-parameter exponential family
\citep{sch2014rmp}, count data \citep{gst2013uhc}, recurrent event
data \citep{holzhauer2018sim}, time-to-event data \citep{hol2018mad}
and variance data \citep{sch2017mvd}.


\section{Application}
\label{sec:application}

In the following the use of \rbest\ is explained for the example
introduced in section \ref{sec:example}. The \rbest\ package
facilitates the (i) prior derivation using MCMC, (ii) parametric
(mixture) approximation of the MAP prior and finally (iii) evaluation
of the clinical trial design.

\subsection{Prior derivation}
\label{sec:app_prior_mcmc}

The statistical models implemented in the package follow the standard
generalized linear regression modelling conventions and are
implemented with the \code{gMAP} function mostly analogous as in the R
\code{glm} command of the \pkg{stats} package. The supported sampling
distributions are normal (with known sampling standard deviation
$\sigma$), binomial and Poisson. These use the canonical link
functions of the identity, logit and $\log$ link, respectively. The
\code{gMAP} function call for the secukinumab trial is:

\begin{Schunk}
\begin{Sinput}
> set.seed(35667)
> map_mcmc <- gMAP(cbind(r, n-r) ~ 1 | study, family=binomial, data=AS,
+                  tau.dist="HalfNormal", tau.prior=1, beta.prior=cbind(0, 2))
\end{Sinput}
\end{Schunk}

The first argument is the formula argument which specifies a
two-column matrix as response for a binary endpoint and contains in
the first column the number of responders $r$ and in the second column
the number of non-responders $n-r$. The response is modeled using an
intercept only here, but covariates can be specified using standard R
formulae syntax. The last element of the formula is the grouping
factor, separated by a vertical bar. The grouping factor defines what
constitutes a trial in the data set. If no grouping factor is
specified, then each row in the input data set is interpreted as a
group. The next argument is the \code{family} argument which specifies
the sampling distribution. It is strongly recommended to use
\code{data} and pass a \code{data.frame} to it where all data for the
model is stored (otherwise the environment will be searched for the
respective columns). Finally, the priors of the model are
specified. As the between-trial heterogeneity parameter $\tau$ is of
particular importance for a meta-analysis, the \code{gMAP} function
allows the user to choose the distributional class for the $\tau$
parameter with the \code{tau.dist} argument whereas the regression
coefficients $\beta$ are restricted to $\normal$ priors. The arguments
\code{tau.prior} and \code{beta.prior} both take a two column argument
with the convention that each row corresponds to the respective
parameter and the columns correspond to respective parameters of the
prior distribution. Whenever a prior distribution for $\tau$ only
needs a single parameter, a vector can be given has well, which is the
case for the $\hnormal$ distribution used here with a standard
deviation of $1$. For the \code{beta.prior}, the first and second
column correspond to the means and standard deviations of the normal
prior distributions, respectively.

Internally the \code{gMAP} function uses \pkg{Stan} via the
\pkg{rstan} package for sampling from the posterior distribution. The
diagnostics of the MCMC sampler as provided by \pkg{Stan} are
automatically inspected and potential issues are reported with a
warning. The \code{gMAP} function returns an \code{S3} object, which
can then be further processed by standard \proglang{R} modeling
functions. The default method

\begin{Schunk}
\begin{Sinput}
> print(map_mcmc)
\end{Sinput}
\begin{Soutput}
Generalized Meta Analytic Predictive Prior Analysis

Call:  gMAP(formula = cbind(r, n - r) ~ 1 | study, family = binomial, 
    data = AS, tau.dist = "HalfNormal", tau.prior = 1, beta.prior = cbind(0, 
        2))

Exchangeability tau strata: 1 
Prediction tau stratum    : 1 
Maximal Rhat              : 1 

Between-trial heterogeneity of tau prediction stratum
  mean     sd   2.5%    50%  97.5% 
0.3830 0.2090 0.0442 0.3550 0.8740 

MAP Prior MCMC sample
  mean     sd   2.5%    50%  97.5% 
0.2590 0.0894 0.1100 0.2480 0.4720 
\end{Soutput}
\end{Schunk}

shows a short summary of the \code{gMAP} analysis. The derived MAP
prior corresponds to the intercept-only model of the fitted
statistical model (relevant when using covariates) and is given on the
response scale by default as opposed to the link scale of $\log$-odds
in this case. More information of the model estimates is available
with the \code{summary} method. Importantly, the \code{plot} method
provides key graphical summaries of the \code{gMAP} analysis:

\begin{Schunk}
\begin{Sinput}
> model_plots <- plot(map_mcmc)
> names(model_plots)
\end{Sinput}
\begin{Soutput}
[1] "densityThetaStar"     "densityThetaStarLink" "forest_model"        
\end{Soutput}
\end{Schunk}

The density estimate plot on the response or link scale of the MAP
prior shows each fitted chain (by default 4 chains are used)
separatley. The overlayed display by chain allows to assess
graphically the convergence of the MCMC analysis, since each chain
must have sampled the same density resulting in very similar
densities. As key diagnostic plot we recommend to inspect the
\code{forest_model} as shown in
Figure~\ref{fig:gMAP_forest_model_plot}. The plot gives a graphical
summary of the MAP analysis and can serve as a fast check for coding
errors.

\begin{Schunk}
\begin{figure}

{\centering \includegraphics[width=\maxwidth]{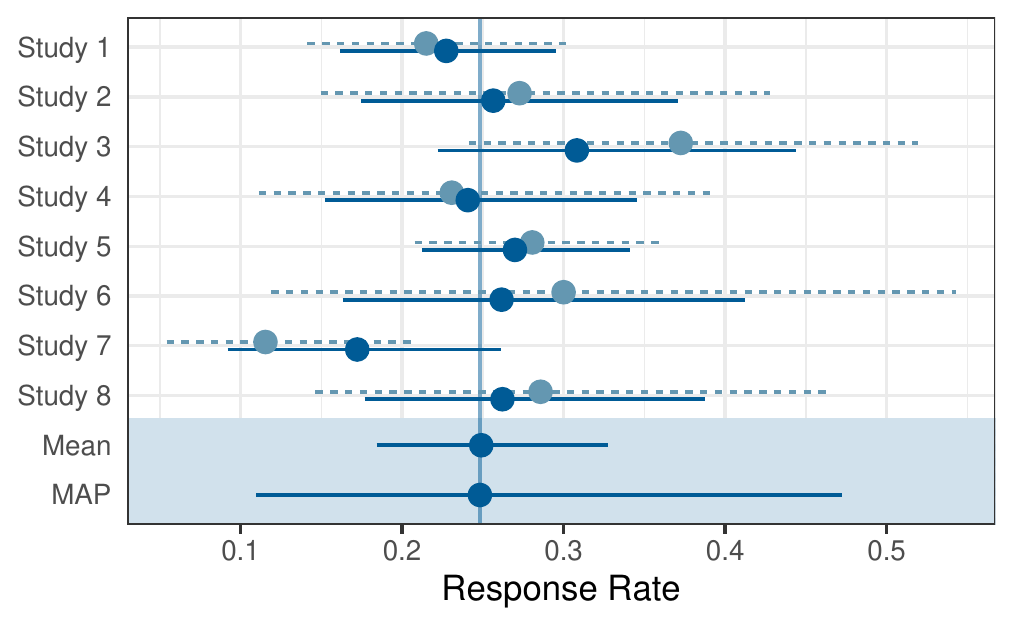} 

}

\caption[Recommended diagnostic plot for \code{gMAP} analyses which is a forest plot augmented with meta-analytic shrinkage estimates per trial]{Recommended diagnostic plot for \code{gMAP} analyses which is a forest plot augmented with meta-analytic shrinkage estimates per trial. Shown are the per-trial point estimates (light dot) and the 95\% frequentist confidence intervals (dashed line) and the model derived median (dark point) and the 95\% credible interval of the meta-analytic model. In addition the model derived typical parameter estimate and the MAP estimate is shown.}\label{fig:gMAP_forest_model_plot}
\end{figure}
\end{Schunk}

\begin{Schunk}
\begin{figure}

{\centering \includegraphics[width=\maxwidth]{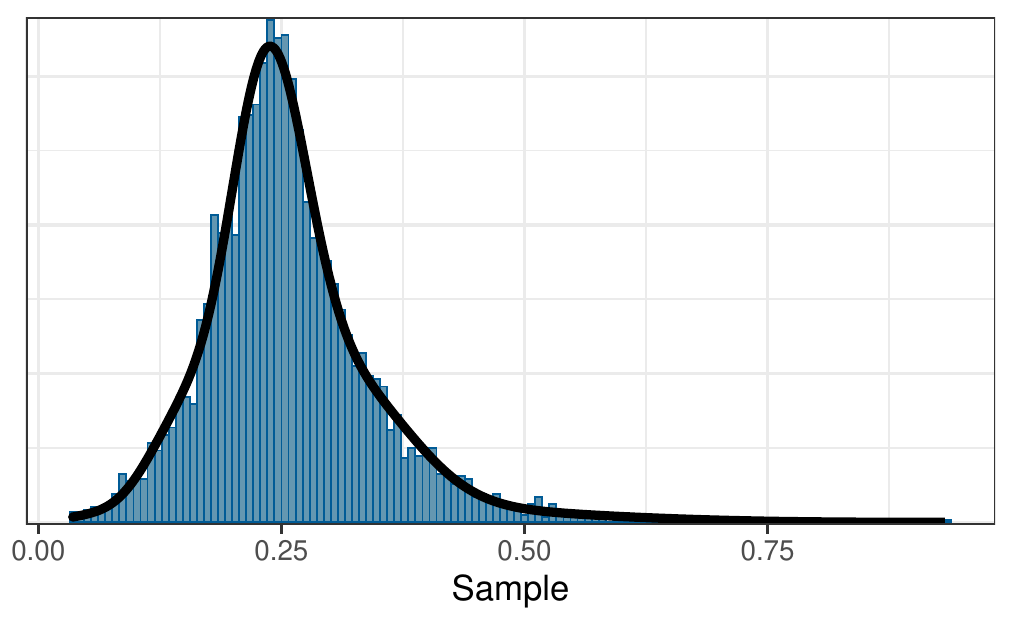} 

}

\caption[Recommended diagnostic plot for parametric approximation of the MCMC represented MAP prior]{Recommended diagnostic plot for parametric approximation of the MCMC represented MAP prior. The plot overlays the MCMC histogram of the MAP prior as obtained from \code{gMAP} and the respective parametric mixture approximation as determined from \code{automixfit} or \code{mixfit}.}\label{fig:mixfit_EM_plot}
\end{figure}
\end{Schunk}

\subsection{Parametric mixture prior derivation}
\label{sec:app_prior_mix}

The MAP prior, represented numerically with a large MCMC simulation
sample, can be approximated with a parametric representation with the
\code{automixfit} function. This function fits a parametric mixture
representation using expectation-maximization (EM) algorithm. When
calling this function with a \code{gMAP} analysis object, the EM
algorithm is run on the MAP prior MCMC values on the response scale
(as opposed to the transformed link scale). The distributional class
of the fitted mixture densities depends on the sampling distribution
of the \code{family} argument. For the \code{gaussian},
\code{binomial} and \code{poisson} family a mixture of normal, beta
and gamma distributions are fitted with the EM algortihm,
respectivley. These choices allow \rbest\ to take advantage of
analytical results as the respective likelihood-prior combinations are
conjugate to one another.

The EM algorithm requires a pre-specified number of components, which
is chosen from \code{automixfit} automatically through the use of
AIC. The function fits parametric mixture models of increasing
complexity with up to four components and then selects the one with
the lowest AIC. The output below shows the log-likelihood results for
the selected mixture model, as well as the mixture components of the
beta mixture. All mixtures are represented in \rbest\ using for each
mixture component $k$ a triplet $(w,a,b)_k$ which correspond to the
weight $w$ of the component, the first standard parameter $a$ and the
second standard parameter $b$ of the respective density. Please refer
to the overview table~\ref{tab:mixtures} for further details.

\begin{Schunk}
\begin{Sinput}
> map <- automixfit(map_mcmc)
> print(map)
\end{Sinput}
\begin{Soutput}
EM for Beta Mixture Model
Log-Likelihood = 4356.121

Univariate beta mixture
Mixture Components:
  comp1      comp2      comp3      comp4     
w  0.4652656  0.2038402  0.1955961  0.1352982
a 31.0317022 21.3507421 10.2829720  2.2980848
b 96.5540272 42.9105627 45.1537087  5.0607874
\end{Soutput}
\end{Schunk}

To consider the quality of the EM fit we recommend to compare the MCMC
sample with the parametric mixture density in a graphical manner. In
Figure~\ref{fig:mixfit_EM_plot} the output of the
\code{plot} method is shown for the generated \code{mix} plot, which
overlays the fitted mixture density marginal with a histogram of the
MCMC sample. In rare cases the response scale
can be inadequate to compare the parametric mixture density
appropriately (for example, if the response rate is very small or very
large):

\begin{Schunk}
\begin{Sinput}
> em_diagnostic <- plot(map)
> print(em_diagnostic$mix)      ## Shown in Figure 2(b)
> print(em_diagnostic$mixdens)  ## Not shown
> 
> em_diagnostic_link <- plot(map, link="logit")
> print(em_diagnostic_link$mix) ## Not shown, same as 2(b) on logit scale
\end{Sinput}
\end{Schunk}

Once the user has derived a parametric mixture representation for the
MAP prior, the \rbest\ package provides additional functions to
further investigate as shown in the overview
Table~\ref{tab:mixtures}. In the following we discuss the key
functions needed in the context of informative prior derivation from
historical data.

As the goal is to reduce the required sample size in a future trial,
the informative MAP prior enables unequal randomization of active vs
control. In the ankylosing spondylitis example a 4:1 randomization
ratio was chosen. The \code{ess} function provides approximations to
the effective sample size of a given prior for various methods. The
effective sample size of the MAP prior gives a rough guide by how much
the sample size can be reduced when using the respective frequentist
power calculation as a reference, for example.

Instead of using the MAP prior directly in a new study, we recommend
to robustify the prior with a weakly-informative component
(Equation~\ref{eqn:rob} of Section~\ref{sec:BESR}) as follows:

\begin{Schunk}
\begin{Sinput}
> map_robust <- robustify(map, weight=0.2, mean=0.5)
\end{Sinput}
\end{Schunk}

\rbest\ offers many functions which facilitate to explore the
implications of an informative prior. For example, the predictive
distribution of the data in a new trial can easily be derived with the
\code{preddist} command.  While the MAP prior is the predictive
distribution of the mean parameter for a new trial and accounts for
parameter uncertainty and between-trial heterogeneity, the predictive
distribution of data accounts in addition for sampling
uncertainty. For beta mixture densities the respective predictive
distribution is the $\betabinomial$ mixture distribution. This can be
used to illustrate possible outcomes in a future trial or to calculate
the Bayes factor of observed data with respect to the prior.

Moreover, \rbest\ provides the \code{postmix} command which updates a
prior with the data as observed and computes analytically the
posterior mixture distribution. For two-sample cases we are often
interested in the difference distribution of two densities
(representing a treatment difference). In \rbest\ the difference
distribution of mixtures of the same class is supported through the
use of the convolution theorem which allows for an accurate
evaluation. The table~\ref{tab:mixtures} summarizes all functions
available for parametric mixtures supported in \rbest. Further details
are available in the help files for each function.

\begin{landscape}
\begin{table}[ht]
\centering
\begin{tabular}{p{3.8cm}|p{8.3cm}|p{8.3cm}}
  Function Name            & Use                                                                & Notes                               \\ \hline
  \code{c(w, a, b)}  & Defines a component of a mixture distribution. Specifies for each component its weight, the
                       first and second standard parameter. &
                                                              Normal
                                                              densities
                                                              use
                                                              mean
                                                              and
                                                              standard
                                                              deviation,
                                                              beta
                                                              densities use
                                                              $\alpha$
                                                              and
                                                              $\beta$,
                                                              gamma densities use shape and rate.
  \\ \hline
  \code{mn2norm} \code{mn2beta} \code{mn2gamma} & Maps to the standard parametrization given mean and number of observations.      &                                     \\ \hline
  \code{ms2beta} \code{ms2gamma}         & Maps to the standard parametrization given mean and standard deviation.          &                                     \\ \hline
  \mbox{\code{mixnorm}} \mbox{\code{mixbeta}} \mbox{\code{mixgamma}} &
                                                                       Create
                                                                       supported
                                                                       base
                                                                       mixture
                                                                       distribution
                                                                       objects.
                                                                                                &
                                                                                                  For normal densities the attribute \code{sigma} sets the known sampling standard deviation. For gamma densities a \code{likelihood} attribute sets the likelihood the density is intended to be used with.                                  \\ \hline


  \code{mixcombine}               & Combines mixture distributions.                 &                                     \\ \hline
  \code{(d/p/q/r)mix}             & Density/distribution/quantile/random
                             number generation functions             &
                                                                       The
                                                                       quantile
                                                             function
                                                                       uses numerical search routines. \\ \hline
  \code{robustify}                & Adds a weakly-informative component
                                    to a mixture.                 &                                     \\ \hline
  \code{(d/p/q/r)mixdiff}                  & Common distribution
                                             functions for the difference of two mixture
                                             distributions.
                                                                                                &
                                                                                                  Uses
                                                                                                  convolution
                                                                                                  theorem
                                                                                                  with
                                                                                                  numerical
                                                                                                  integration while
                                                                                                  for
                                                                                                  normal
                                                                                                  mixtures
                                                                                                  analytically
                                                                                                  exact
                                                                                                  results
                                                                                                  are used.
  \\ \hline
  \code{ess}                      & Calculates the effective sample
                                    size of a mixture distribution.
                                                                                                &
                                                                                                  Methods
                                                                                                  implemented
                                                                                                  are
                                                                                                  \code{elir}
                                                                                                  (default),
                                                                                                  \code{moment}
                                                                                                  and
  \code{morita}, please refer to appendix~\ref{app:ess}. \\ \hline
  \code{postmix}                  & Calculates posterior mixture
                                    distribution.       & Uses analytically
                                                     exact results.                                   \\ \hline
  \code{preddist}                 & Returns the predictive mixture distribution.                                           &                                     \\ \hline
  \code{decision1S_boundary} \code{decision2S_boundary} & Calculates
                                                          critical
                                                          decision
                                                          boundary. &
                                                                      Please
                                                                      refer to
                                                             appendix~\ref{app:prior_eval}. \\ \hline
  
  \code{oc1S} \code{oc2S} & Calculates conditional power. &
                                                             Please refer to
                                                             appendix~\ref{app:prior_eval}. \\ \hline
  \code{pos1S} \code{pos2S} & Calculates probability of success. &
                                                                  Please refer to
                                                             appendix~\ref{app:prior_eval}. \\ \hline
  
  \code{summary}                  & Produces descriptive statistics of mixture.                        &                                     \\ \hline
  \code{plot}                  & Produces diagnostic plots.                             &                                     \\
\end{tabular}
\caption{Overview on functions in \rbest\ defined for parametric mixtures.}
\label{tab:mixtures}
\end{table}
\end{landscape}

\subsection{Trial design evaluation}
\label{sec:app_design_eval}

Once a parametric mixture MAP prior has been derived, it is crucial to
understand its influence on the statistical analysis of a clinical
trial. In the context of the historical control example the goal is
commonly to reduce the control group sample size while maintaining
adequate power for trial success under the alternative hypothesis
which assumes some true effect size. The main focus here is the
evaluation of the (frequentist) operating characteristics of the trial
design with respect to achieving trial success.

\rbest\ follows a modular, step-wise approach for design
evaluation. First, a success criterion is defined, then the design of
the trial is specified and finally the desired evaluation of the trial
can be conducted for possible scenarios of interest. The success
criterion supported are restricted to one-sided criterion's which are
referred to as decision functions. These can be set up for one-sample
(\code{decision1S}) and two-sample (\code{decision2S}) situations. In
\rbest\ the decision functions can be set up to require that multiple
critical probability thresholds and quantiles for the difference
distribution have to be met in the two-sample case while these
thresholds are applied directly to the posterior density in the
one-sample case. This enables evaluation of so-called dual criterion
designs \citep{Roychoudhury2018} which evaluate a statistical
confidence criterion (high confidence in a positive/negative
difference) and a minimally observed treatment difference (observed
median difference being greater/smaller than some value). For the
ankylosing spondylitis trial, success was declared whenever the
posterior treatment difference is positive with a probability which
exceeds $95\%$. In \rbest\ this is expressed as:

\begin{Schunk}
\begin{Sinput}
> # decision function as used in the trial
> success <- decision2S(0.95, 0, lower.tail=FALSE)
> # an alternative which demands in addition to see at least
> # a median difference of at least 10%
> success_dual <- decision2S(c(0.95, 0.5), c(0, 0.1), lower.tail=FALSE)
\end{Sinput}
\end{Schunk}

The returned object represents the decision function and is a binary
function. It takes as arguments two mixture densities
(those will be the mixture posteriors) and returns \code{1} whenever
the condition for success on the difference distribution of the two
mixture densities is fulfilled and \code{0} otherwise. Optionally, the
success criterion allows for transformation of the input mixture
distributions prior to forming the difference distribution using the
\code{link} argument. The $\log$ and $\logit$ link enable relative
risk and $\log$ odds-ratio success criterion to be evaluated with
\rbest.

The next step is to define the design of the trial using the operating
characeristics function \code{oc2S} (or \code{oc1S} for a one-sample
case). The design of the trial includes the priors for each arm, the
total sample size per arm and the decision function for trial
success. To evaluate, for example, the impact a robust MAP prior has
on the frequentist operating characteristics compared to the
respective non-robust design, one may use:

\begin{Schunk}
\begin{Sinput}
> ## prior used for treatment in trial
> treat_prior <- mixbeta(c(1, 1/2, 1))
> ## explore different trial designs
> design_nonrobust <- oc2S(treat_prior, map       , 24, 6, success)
> design_robust    <- oc2S(treat_prior, map_robust, 24, 6, success)
\end{Sinput}
\end{Schunk}

The \code{oc2S} function returns a binary function which is finally
used to calculate the frequency of trial success as a function of an
assumed truth for each arm. For the binomial sampling distribution,
the function takes assumed true response rates $\theta_1$ and
$\theta_2$. Whenever these two are set to the same numerical value
$\theta=\theta_1=\theta_2$, the scenario of no treatment difference is
evaluated which would be referred to a type-I error in respective
Frequentist trial analysis. Setting $\theta_2$ to a plausible control
response rate and varying $\theta_1$ as a function of the difference
$\delta = \theta_1-\theta_2$ gives the desired power of the trial. The
functions take vectors of equal length as arguments so that we can
evaluate the type-I over large parameter ranges as:

\begin{Schunk}
\begin{Sinput}
> theta  <- c(0.25, 0.5, 0.75)
> round(design_nonrobust(theta, theta), 3)
\end{Sinput}
\begin{Soutput}
[1] 0.020 0.320 0.598
\end{Soutput}
\begin{Sinput}
> round(design_robust(theta, theta), 3)
\end{Sinput}
\begin{Soutput}
[1] 0.018 0.190 0.173
\end{Soutput}
\end{Schunk}

Here we see that the Bayesian design with informative priors does not
control the type-I error and that the error rate depends on the actual
parameter value. However, the $97.5$\% quantile of the MAP prior is
approximatley $0.48$ such that response rates greater than this would
be very unusual. Given that \rbest\ is fast and accurate for these
calculations, it is recommended to use graphical plots in addition to
tables to visualize the (error) rates of interest as demonstrated in
the vignettes of \rbest.

In addition to the operating characteristics functions \rbest\ also
provides respective functions to evaluate the probability of success
with \code{pos2S} and \code{pos1S}. These functions allow to account
for uncertainty in the assumed true parameter values. They
\code{pos2S} (\code{pos1S}) require as arguments the same trial design
specification arguments as the \code{oc2S} (\code{oc1S})
functions. The returned functions take as arguments mixture densities
which represent uncertainty in the respective parameter of each arm.

Internally, all trial design calculations use analytical results
wherever possible. This makes \rbest\ very accurate and fast in
evaluating the design properties of trials. A key quantitiy calculated
is the critical decision boundary determined by the success criterion
and the trial design. At the critical outcome boundary the success
criterion changes its value between $0$ and $1$. While in the one-sample
case this corresponds to a single value, it is a function of the
outcome in the second sample in the two-sample case. As the decision
boundary can be useful for other applications, the boundary can be
obtained with the functions \code{decision2S_boundary}
(\code{decision1S_boundary}). These are also useful when communicating
various data scenarios and their respective decisions to
non-statisticians like clinical teams. For more details please refer
to the appendix.





\section{Summary}
\label{sec:summary}

In this paper, we introduced the \rbest\ package which implements the
MAP approach via MCMC sampling algorithms for a number of common
sampling and prior distributions. Incorporating historical data in
clinical studies has various advantages, such as reducing the number
of subjects randomized to a control arm or getting more precise
information for decision making. Incorporation of historical data
should lead to more ethical and efficient clinical trials.

The MAP approach is a hierarchical modeling method allowing
heterogeneity between historical trials, which can incorporate
historical data in a meta analytic framework. The \rbest\ package
allows for easier implementation of MAP priors. After selection of
appropriate historical information, \rbest\ facilitates the prior
derivation using MCMC, the parametric (mixture) approximation of the
MAP prior and finally the evaluation of the clinical trial design.


\section*{Acknowledgments}

We would like to thank Beat Neuenschwander for his detailled comments
on the manuscript and valuable input during development of \rbest. In
addition we thank for their input Baldur Magnusson, Satrajit
Roychoudhury, Simon Wandel and Björn Holzhauer.


\nocite{assertthat, mvtnorm1, mvtnorm2, Formula, checkmate, bayesplot1,
  bayesplot2, ggplot2, dplyr, BH, Rcpp1, Rcpp2, Rcpp3, RcppEigen}

\bibliography{refs,synthesis,refs_intro,Rpackages}


\newpage

\begin{appendix}

\section{Parametric mixture inference}
\label{app:mixture_fit}

In \rbest\ the expectation maximization (EM) algorithm is used to find
parametric mixture approximations to the numerical representation of
the MAP prior. Thus, we consider in this section as data $\bm{Y}$ the
MCMC sample representing the MAP prior (or any other MCMC
sample). Direct application of maximum likelihood is numerically
problematic, since the $\log$-likelihood for a mixture prior,
$\log p(\bm{Y}|\bm{w},\bm{a},\bm{b}) = \sum_{i=1}^N \log \sum_{k=1}^K
w_k \, p_k(Y_i|a_k, b_k)$, involves the sum over the $\log$ of the
component densities $w_k \, p_k(Y_i|a_k, b_k)$. The inner summation is
on the natural scale and is required as we do not know the component
$k$ which a data item $Y_i$ is a member of. However, extending the
observed data ($\bm{Y}$), also referred to as incomplete data, to the
so-called complete data ($\bm{Y}$,$\bm{Z}$) leads to a numerically
stable problem. The unobserved data $\bm{Z}$ is defined as the latent
component indicator such that $Z_{i,k}$ is $1$ whenever data item $i$
is part of component $k$ and $0$ otherwise. The extended problem is
related to the original through marginalization,
$p(\bm{Y}|\bm{w},\bm{a},\bm{b}) =
E_{\bm{Z}}[p(\bm{Y},\bm{Z}|\bm{w},\bm{a},\bm{b})]$. However, the
extended $\log$-likelihood factorizes in the usual way

\begin{equation*}
  \log p(\bm{Y},\bm{Z}|\bm{w},\bm{a},\bm{b}) = \sum_{i=1}^N
 \sum_{k=1}^K Z_{i,k} \left[ \log(w_k) + \log p_k(Y_i|a_k, b_k)
 \right] .
\end{equation*}

The EM algorithm begins with a fixed number of mixture components
$K$ and an initial guess of all parameters. The initial guess of the
parameters is achieved with the k nearest neighbors (knn) algorithm in
\rbest\ with the exception of the normal mixture case as detailed
below. The parameter vector $(\bm{w},\bm{a},\bm{b})$ is then updated
iteratively. The $n$th iteration of the EM consists of first
performing the \emph{expectation} (E) step and then the
\emph{maximization} (M) step which updates the parameter vector for
the next iteration.  These EM steps are then repeated until
convergence to a maximum of the $\log$-likelihood. While it is
guaranteed that in each iteration the $\log$-likelihood is always
increased, the EM algorithm may only find a local rather than a global
extremum.

\paragraph{E-step}
The E-step calculates the posterior probability for the latent indicators
$p(\bm{Z}|\bm{Y},\bm{w},\bm{a},\bm{b})$ in order to
compute the expected posterior mean weights
\begin{equation*}
  E[Z_{i,k}] = \frac{w_k \, p(Y_i|a_k,b_k)}{\sum_k w_k \,
 p(Y_i|a_k,b_k)} = \gamma(Z_{i,k}) .
\end{equation*}
The expression $\gamma(Z_{i,k})$ is often referred to as the
responsibility of mixture component $k$ for data item $i$. The overall
responsibility of mixture component $k$
\begin{equation*}
  N_k = \sum_{i=1}^N \gamma(Z_{i,k})
\end{equation*}
can be interpreted as the number of data items belonging to mixture
component $k$.
Finally, the E-step computes the expectation of the complete
$\log$-likelihood with respect to $Z$ conditional on the parameter
vector of the current $n$th iteration,
\begin{align*}
  E_{Z|\bm{w}^n,\bm{a}^n,\bm{b}^n}[\log p(\bm{Y},\bm{Z}|\bm{w},\bm{a},\bm{b})] &= \sum_{i=1}^N
 \sum_{k=1}^K \gamma(Z_{i,k}) \left[ \log(w_k) + \log p_k(Y_i|a_k,
                                                                                 b_k) \right] \\
  &= Q(\bm{w},\bm{a},\bm{b}|\bm{w}^n,\bm{a}^n,\bm{b}^n)
\end{align*}

\paragraph{M-step}
The M-step then proceeds by finding a new parameter vector through
maximization
\begin{equation*}
  (\bm{w}^{n+1},\bm{a}^{n+1},\bm{b}^{n+1}) = \argmax_{\bm{w},\bm{a},\bm{b}}
  Q(\bm{w},\bm{a},\bm{b}|\bm{w}^n,\bm{a}^n,\bm{b}^n) .
\end{equation*}
The updated weights are constrained to sum to one. Maximization with
this constraint is achieved through Lagrange multipliers and leads for
the updated weights to the solution
\begin{equation*}
  w_k^{n+1} = \frac{N_k}{N} .
\end{equation*}
To find the maximum with respect to the parameters of each component
$k$, we take the gradient $(\partial_{a_k}, \partial_{b_k})$ of
$Q(\bm{w},\bm{a},\bm{b}|\bm{w}^n,\bm{a}^n,\bm{b}^n)$ and equate these
to zero.

\paragraph{Normal mixtures}
For normal mixtures \rbest\ implements internally a multi-variate
normal EM, but only exposes the uni-variate functionality at the
moment. Empirically we observed that the normal EM algorithm is easily
trapped into local extrema which is caused by the commonly heavy
tailed distributions of MAP priors. For this reason, we initialize the
normal EM with the result of a Student-t EM procedure as described in
\cite{Peel2000}. The Student-t EM is itself initialized with k nearest
neighbors. The maximization equations can be solved analytically in
the normal mixture case,
\begin{align*}
  \mu_k &= \frac{1}{N_k} \, \sum_{i=1}^{N} \gamma(Z_{i,k}) \, Y_i \\
  \Sigma_k &= \frac{1}{N_k} \, \sum_{i=1}^{N}  \gamma(Z_{i,k}) \,
             (Y_i-\mu_k)\, (Y_i-\mu_k)' .
\end{align*}
The analytical solution is a weighted mean and covariance estimate
with the weight for each data item $Y_i$ equal to
$\gamma(Z_{i,k}) /N_k$.

\paragraph{Beta mixtures}
For beta mixture distributions we are lead to the joint equation
system of \citep[see also][]{Ma2009}
\begin{align*}
  \psi(a_k) - \psi(a_k + b_k) &= \frac{1}{N_k} \, \sum_{i=1}^N \gamma(Z_{i,k}) \,
  \log(Y_i) \\
  \psi(b_k) - \psi(a_k + b_k) &= \frac{1}{N_k} \, \sum_{i=1}^N \gamma(Z_{i,k}) \,
  \log(1-Y_i).
\end{align*}
Here $\psi(x)$ is defined as $\partial_x \log(\Gamma(x))$. This
equation system is solved simultaneously for $a_k$ and $b_k$ through
numerical minimization.

\paragraph{Gamma mixtures}
With gamma mixtures the algebraic equation system
\begin{align*}
  \psi(a_k) - \log(b_k) &= \frac{1}{N_k} \, \sum_{i=1}^N \gamma(Y_{i,k})
                                        \, \log(Y_i) \\
  \frac{a_k}{b_k} &= \frac{1}{N_k} \, \sum_{i=1}^N \gamma(Z_{i,k}) \, Y_i
\end{align*}
is obtained. This system can be reduced to a single equation, which is
again solved through numerical minimization.

\section{Effective sample size}
\label{app:ess}

The effective sample size is an approximate measure for the number of
observations a prior is equivalent to. In the setting of conjugate
likelihood-prior pairs without mixtures the standard parameters can be
cast into an effective sample size. However, this is not the case for
mixture priors and \rbest\ implements three approaches:

\begin{description}
\item[elir] The expected local information ratio has been proposed in
  \cite{Neuenschwander2019B} and is a predictively consistent
  effective sample size measure. The predictive consistency requires
  that the effective sample size of the prior is equal to the average
  effective sample size of the respective posterior of this prior
  after simulation of $m$ samples from the predictive prior
  distribution and subtracting $m$ from the averaged posterior
  effective sample size. The method is neither liberal nor
  conservative and is the default method in \rbest.
\item[morita] The method from \cite{mor2008des} evaluates the
  curvature of the prior at a reference point (mode, median or mode)
  and compares this against the expected curvature of a posterior of a
  variance inflated prior which is updated with $m$ samples from the
  prior predictive of the prior. The $m$ is chosen to minimize the
  difference in curvature. Since the Morita method evaluates the prior
  at a single point, the approach can be sensitive to the curvature at
  this point and has been observed to report relatively liberal
  effective sample sizes when used with mixture priors.
\item[moment] The moment based approach matches the mean and the
  variance of a given prior to its respective canonical prior from
  which the effective sample size can be obtained directly from the
  standard parameters. This approach has been found empirically to
  report very conservative (low) effective sample sizes.
\end{description}

The key expressions involved in the effective sample size calculations
is the prior information
\begin{equation*}
  i(p(\theta)) = - \pd[2]{\log(p(\theta))}{\theta}
\end{equation*}
and the Fisher information
\begin{equation*}
  i_F(\theta) = E_{Y_1|\theta}\left[i(p(Y_1|\theta))\right] =
  -E_{Y_1|\theta}\left[\pd[2]{\log(p(Y_1|\theta))}{\theta} )\right] .
\end{equation*}
The Fisher information is derived from the sampling distribution, but
the second derivative of the $\log$ mixture prior density with $K$
components, defined as
\begin{equation*}
  p(\theta|\bm{w},\bm{a},\bm{b}) = \sum_{k=1}^K w_k\, p_k(\theta|a_k,b_k),
\end{equation*}
needs some considerations for a numerically stable evaluation. It is
key to evaluate the $\log$ density instead of the density on the
natural scale whenever possible. We will in the following suppress the
weights and standard parameters of $p(\theta)$ and $p_k(\theta)$
for simplicity. Using the equality
$\pd{p(\theta)}{\theta} = p(\theta) \, \pd{\log(p(\theta))}{\theta}$
one finds that the prior information for a mixture is

\begin{align}
  i(p(\theta)) =& \, \frac{1}{p(\theta)^{2}} \, \left[ \sum_{k=1}^{K} w_k \,
    p_k(\theta) \, \pd{\log p_k(\theta)}{\theta} \right]^2 \nonumber \\
  &-\frac{1}{p(\theta)} \, \sum_{k=1}^{K} w_k \,  p_k(\theta) \, \left[ \left(\pd{\log p_k(\theta)}{\theta}\right)^{2} + \pd[2]{\log p_k(\theta)}{\theta} \right].
\end{align}

The table~\ref{tab:densities} lists all the main expressions required
for the supported conjugate likelihood-prior pairs in \rbest.

\begin{table}[ht]
\centering
\begin{tabular}{l|l|l|l}
  Sampling distribution & Fisher information & Prior density & Prior
                                                               information
  \\
  \multicolumn{1}{c|}{$p(Y|\theta)$} & \multicolumn{1}{c|}{$i_F(\theta)$} & \multicolumn{1}{c|}{$p_k(\theta|a,b)$} & \multicolumn{1}{c}{$i(p_k(\theta))$} \\
  \hline
  $\normal(Y|\theta, \sigma^2)$ & $\sigma^{-2}$ & $\normal(\theta|m, s^2)$ &
                                                                       $s^{-2}$
  \\
  $\dbinomial(Y|\theta, n)$ & $\frac{n}{\theta \, (1-\theta)}$ &
                                                                 $\dbeta(\theta|a,b)$
                                                             &
                                                               $\frac{a-1}{\theta^2}
                                                               +
                                                               \frac{b-1}{(1-\theta)^2}$
  \\
$\dpois(Y|\theta)$ & $\theta^{-1}$ & $\dgamma(\theta|a,b)$ &
                                                                  $\frac{a-1}{\theta^2}$
   \\
 $\dexp(Y|\theta)^\dagger$ & $\theta^{-2}$ & $\dgamma(\theta|a,b)$ & $\frac{a-1}{\theta^2}$
\end{tabular}
\caption{Overview on supported conjugate likelihood-prior pairs
  supported in \rbest. $^\dagger$Note that for the exponential
  sampling distribution only the effective sample size calculations
  are supported as of \rbest\ 1.5-4.}
\label{tab:densities}
\end{table}

\section{Informative prior evaluation}
\label{app:prior_eval}

In \rbest\ one-sided decision functions with multiple criteria are
supported for one and two sample cases. The decision functions are
indicator functions through thresholding density distributions such that
critical quantiles must exceed predefined probability
thresholds. Denoting with $H(x)$ the Heaviside step function, which is
$0$ for $x \leq 0$ and $1$ otherwise, the decision functions are
defined as
\begin{align*}
  D(p(\theta)) &= \prod_i \, H(P(\theta \leq q_{i,c}) - p_{i,c}) &
                                                                 \mbox{one
                                                                 sample,} \\
  D(p_1(\theta_1),p_2(\theta_2)) &= \prod_i \, H(P(\theta_1 - \theta_2 \leq q_{i,c}) - p_{i,c}) &
                                                                                                \mbox{two
                                                                                                sample.}
\end{align*}
In the two-sample case the difference distribution of $p_1(\theta_1)$
and $p_2(\theta_2)$ is thresholded.

\paragraph{Critical decision boundary} With the design of a trial the
priors and the sample size per sample is defined and these determine
the critical decision boundary of the decision function,
\begin{align*}
  D_c &= \sup_y\{D(p(\theta|y)) = 1\} = y_c &
                                              \mbox{one
                                              sample,} \\
  D_{c,1}(y_2) &=  \sup_{y_1}\{D(p_1(\theta_1|y_1), p_2(\theta_2|y_2)) = 1\} &
                                                                               \mbox{two
                                                                               sample.}
\end{align*}
While for the one sample cases the decision boundary is a single
value, $D_c=y_c$, the decision boundary is a function of the outcome
in one of the samples (by convention the second sample),
$D_{c,1}(y_2)$.

\paragraph{Conditional power} The critical decision boundary is used
in \rbest\ to simplify the conditional power calculation in the
following manner,
\begin{align*}
  \cp(\theta) &= \int D(p(\theta'|y)) \, p(y|\theta) \, dy
                  = \int_{-\infty}^{y_c} p(y|\theta) \, dy &
  \\
                &= P(y \leq y_{c}|\theta) & \mbox{one
                                              sample,} \\
  \cp(\theta_1,\theta_2) &= \iint D(p_1(\theta_1'|y_1),p_2(\theta_2'|y_2)) \, p_1(y_1|\theta_1) \, 
                           p_2(y_2|\theta_2)  \, dy_1\,dy_2 & \\
  &= \int P_1(y_1 \leq D_{c,1}(y_2)|\theta_1) \, p_2(y_2|\theta_2) \, dy_2 &
                                                                \mbox{two
                                                                sample.}
\end{align*}
Therefore, the conditional power simplifies in the one sample case to
evaluation of the cumulative density function corresponding to the
sampling distribution and in the two sample case the integration is
simplified to a one dimensional integral instead of a two dimensional
one. For the case of a binomial sampling distribution the calculations
are carried out exactly. With normal and Poisson endpoints the
respective integrals are evaluated with quadrature integration on a
domain which corresponds to $1-\epsilon$ ($\epsilon = 10^{-6}$ by default) of
probability mass.

\paragraph{Probability of success} The probability of success as
defined in Equation~\ref{eqn:pos} of Section~\ref{sec:BESA.POS}
results in a double (triple) integral for the one (two) sample
case. To simplify this, the prior predictive distribution of the prior
$p(\theta)$,
\begin{equation*}
  p(y) = \int p(y|\theta) \, p(\theta) \, d\theta,
\end{equation*}
is used. The prior predictive distribution is available in analytic
form and allows to re-arrange the evaluation of the
Equation~\ref{eqn:pos} as
\begin{equation*}
  \pos = \int \cp(\theta) \, p(\theta) \, d\theta = \iint D(p(\theta'|y)) \,
  p(y|\theta) \, p(\theta) \, d\theta \, dy = \int D(p(\theta'|y)) \,
  p(y) \, dy.
\end{equation*}
This re-arrangement holds for the two sample case respectively. This
leads to the same calculations as previously for the operating
characteristics with the only difference in that the sampling
distribution, $p(y|\theta)$, is replaced by the predictive
distribution of the prior, $p(y)$. For the two sample case the
integration is performed using numerical integration while for the one
sample case the cumulative distribution function of the predictive is
evaluated.




\end{appendix}


\end{document}